\documentclass[12pt]{elsart}
\usepackage{amsmath,amssymb}
\usepackage{graphicx,psfrag,epsfig,citesort}
\usepackage{pstricks}
\usepackage{pst-node}
\usepackage{epsf}
\usepackage{dsfont}
\allowdisplaybreaks[1]

\begin{document}

\def\eq{\begin{eqnarray}}
\def\en{\end{eqnarray}}

\def\query#1{\marginpar{\begin{flushleft}\footnotesize#1\end{flushleft}}}%

\runauthor{D\"oring and Mei{\ss}ner}
\renewcommand{\theequation}{\arabic{equation}}
\begin{frontmatter}

\title{\Large\bf Kaon-nucleon scattering lengths from kaonic 
deuterium experiments revisited}

\author[Bonn]{M. D\"oring,}
\author[Bonn,Juelich]{U.-G. Mei{\ss}ner}

\address[Bonn]
{Universit\"{a}t Bonn, Helmholtz-Institut f\"{u}r
Strahlen- und Kernphysik (Theorie) and Bethe Center for Theoretical 
Physics, D-53115 Bonn, Germany}

\address[Juelich]
{Forschungszentrum J\"{u}lich, Institut f\" {u}r Kernphysik, Institute for 
Advanced Simulation and J\"ulich Center for Hadron Physics,
D-52425 J\"{u}lich, Germany}

\begin{abstract}
We analyse the impact of the recent measurement of kaonic hydrogen  X rays by
the SIDDHARTA collaboration on the allowed ranges for  the kaon-deuteron
scattering length in the framework of non-relativistic effective field theory.
Based on data from $\bar KN$ scattering only, we predict the kaon-deuteron scattering length
$A_{Kd}= (-1.46 + i\ 1.08)~{\rm fm}$, with an estimated uncertainty of about 25\% 
in both the real and the imaginary part.
\end{abstract}

\end{frontmatter}

{\bf 1.}
Recently, the SIDDHARTA collaboration at LNF-INFN has performed a measurement
of the energy level shift ($\epsilon_{1s}$) and width ($\Gamma_{1s}$) of the
kaonic hydrogen ground-state \cite{Bazzi:2011zj}
\eq
\epsilon_{1s}&=&-283\pm 36~\mbox{(stat)}\pm 6~\mbox{(syst) eV}\, ,
\nonumber\\[2mm]
\Gamma_{1s}&=&\,\,\,\,541\pm 89~\mbox{(stat)}\pm 22~\mbox{(syst) eV}~,
\en
which allows to extract the fundamental antikaon-proton ($K^-p$) scattering
length based on an improved Deser-type formula developed in~\cite{Raha1}. This
measurement resolved the long-standing puzzle of the discrepancy between the
earlier DEAR ~\cite{Beer}  and the less accurate  KpX experiment at
KEK~\cite{KEK}. The DEAR data have been puzzling the community for a long time.
As first pointed out in Ref.~\cite{Raha1}, the energy shift and width of kaonic
hydrogen measured by DEAR is incompatible with the predicted values taking the
underlying $\bar KN$ scattering lengths from scattering data only. This issue
was studied and exposed in more detail in a series of papers by various groups,
see e.g.  Refs.~\cite{OPV,BNW,Oller:2006jw,Borasoy:2006sr}. In addition, based
on the framework of non-relativistic effective field theory (for a recent
comprehensive  review with many  applications to hadronic atoms, see
Ref.~\cite{Gasser:2007zt}), it was shown in~\cite{Meissner:2006gx} that with
the DEAR central values for the kaonic hydrogen ground-state energy and width,
a solution for the isoscalar ($a_0$) and the isovector ($a_1$) kaon-nucleon
scattering lengths exists only in a very restricted domain of input values for
the kaon-deuteron scattering  length. Consequently, it was concluded that the
anticipated measurement of kaonic deuterium by the SIDDHARTA collaboration 
\cite{Cargnelli:2005pd,Marton:2006bq}  would pose stringent constraints on the
kaon-deuteron interaction at low energies. Presently, an upgrade of the
detector at LNF-INFN, called SIDDHARTA2, is being considered to perform
measurements of  X rays in kaonic deuterium in 2012
\cite{Curceanu(Petrascu):2011zz}. It is therefore of high interest to reanalyse
the predictions for kaonic deuterium in the light of the new kaonic hydrogen
measurements. This is exactly what will be done in this Letter. 

\medskip

\noindent
{\bf 2.} 
To reanalyse the predictions for kaonic deuterium, the elementary 
antikaon-nucleon ($\bar KN$) scattering lengths $a_{\bar KN}$ have to be 
related to the $K^-d$ scattering length $A_{Kd}$. As the $a_{\bar KN}$ are 
comparable in size to the average distance of the nucleons in the deuteron, 
the multi-scattering series is non-perturbative and needs to be resummed.  The
resummed fixed-center-approximation (FCA) to the $Kd$ problem has been
formulated in Ref.~\cite{Kamalov:2000iy}, 
see also Ref.~\cite{Meissner:2006gx}. A three-body calculation beyond the FCA
has been performed, e.g., in Ref.~\cite{Bahaoui:2003xb} where the loop momentum
integration is retained. This allows to take account of the strong energy 
dependence of the elementary scattering processes near threshold.
Recoil corrections  beyond the
FCA can in principle be included in a controlled way in a systematic expansion
in integer and half-integer powers of the parameter  $\xi=M_K/m_N \simeq 1/2$
with $M_K \ (m_N)$ the kaon (nucleon) mass.
For double scattering this has been shown in  Ref.~\cite{Baru:2009tx}, see also
Ref.~\cite{Baru:2004kw}. A consistent inclusion of these corrections in the
multiple scattering series is still  an open issue. Here, we will rely on the
resummed FCA approximation without  recoil corrections (except for trivial
kinematical factors) of  Refs.~\cite{Kamalov:2000iy,Meissner:2006gx}, 
\begin{equation}
\label{eq:ratio-Kamalov}
\hat a_{Kd}(r)=\frac{\tilde a_p+\tilde a_n
+(2\,\tilde a_p\,\tilde a_n-b_x^2)/r-2\,b_x^2\,\tilde a_n/r^2}
{1-\tilde a_p\,\tilde a_n/r^2+b_x^2\,\tilde a_n/r^3}+\delta \hat a_{Kd} ~,
\end{equation}
with $b_x^2=\tilde a_x^2/(1+\tilde a_u/r)$. The $K^-d$ scattering length $A_{Kd}$
is obtained from $\hat a_{Kd}$ via a folding with the deuteron wave function
and the $\tilde a$ on the right-hand side are related  to the elementary
scattering lengths $a_{\bar KN}$~\cite{Kamalov:2000iy,Meissner:2006gx}.  Here,
we employ for simplicity the CD-Bonn potential  for the
$S$- and $D$-wave parts of the wave function. In Eq.~(\ref{eq:ratio-Kamalov}), 
the indices $p,\,n,\,x$, and $u$ refer to the  processes $K^-p\to K^-p$,
$K^-n\to K^-n$, $K^-p\to \bar K^0 n$, and  $\bar K^0 n\to \bar K^0 n$, in
order. The quantity $\delta \hat a_{Kd}$ indicates a genuine three-body piece
which is not determined but argued to be  small in Ref.~\cite{Meissner:2006gx},
and thus $\delta \hat a_{Kd}=0$ in the present study. This issue deserves
further investigations in the future.

To determine the influence of the isoscalar and isovector scattering lengths,
$a_0$ and $a_1$, on the $Kd$ scattering length, one has to relate them to the 
elementary processes of Eq.~(\ref{eq:ratio-Kamalov}). This has been achieved in
Ref.~\cite{Raha1} by means of effective field theory up to next-to-leading
order in isospin breaking. There, it has been shown that the large leading
order isospin correction is provided by the unitary cusp. By resumming neutral
kaon loops, the elementary scattering lengths in the particle basis can be
related to $a_0$ and $a_1$ as shown in detail in Ref.~\cite{Raha1}. 
The result for $a_p$ as a function of $a_0,a_1$ can be rewritten as
\begin{equation}
a_p = \frac{(a_0+a_1)/2+q_0a_0a_1}{1+{q_0}(a_0+a_1)/2}\, .
\label{aorew}
\end{equation}
This condition, together with the requirement from unitarity, ${\rm
Im}\,a_I\geq 0$, $i=0,1$, leads to a restriction for the possible values  of
$a_0,\,a_1$~\cite{Meissner:2006gx} in form of circles in the complex $a_I$ 
planes, i.e. values of $a_I$ inside these circles are excluded.  The circle is
given by 
\eq
a_I^{\rm lim}&=&C+R\exp(i\phi),\quad \phi,R\in \mathds{R},\nonumber \\
\frac{1}{R}&=&-4\,q_0\,{\rm Im}\,\frac{1}{1+a_{p}\,q_0},\quad
C=-\frac{1}{q_0}+i\,R~,
\label{circle}
\en
for both $I=0$ and $I=1$. In Eqs.~(\ref{aorew}) and (\ref{circle}), the
parameter $q_0$ determines the strength of the cusp, $q_0=\sqrt{2\mu_0\Delta}$,
where $\mu_0$ is the reduced mass of the $\bar K^0$ and the $n$, and
$\Delta=m_n+M_{\bar K^0}-m_p-M_{K}$ as derived in Ref.~\cite{Raha1}.

Within this framework, we can now study the $K^-d$ scattering length $A_{Kd}$, 
using the constraints provided by $a_p$.  The ground-state energy shift and width of kaonic
hydrogen can be related  to the $K^-p$ scattering length $a_p$ at
next-to-leading order in isospin breaking as derived in Ref.~\cite{Raha1},
\begin{equation}
\epsilon_{1s}-\frac{i}{2}\,\Gamma_{1s} = -2\alpha^3\mu_c^2\, a_p
\biggl(1-2\alpha\mu_c(\ln\alpha-1)\,a_p\biggr)
\label{apnlo}
\end{equation}
where $\mu_c$ is the reduced mass of the $K^-$ and the $p$. In
Tab.~\ref{tab:ap} values for $a_p$ extracted from different  experiments are
shown.
\begin{table}[t!]
\begin{center}
\noindent\begin{tabular}{|l|l|}
\hline \hline 
$a_p$ [fm]	& Experiment\\
\hline
$-0.82+i\,0.64$	& KpX~\cite{KEK} \\
$-0.48+i\,0.35$	& DEAR~\cite{Beer} \\
$-0.66+i\,0.81$	& SIDDHARTA~\cite{Bazzi:2011zj} \\
$-0.85+i\,0.78$	& Average SIDDHARTA~\cite{Bazzi:2011zj} \&
scattering~\cite{Borasoy:2006sr} \\
\hline \hline
\end{tabular}
\bigskip
\caption{Values of the $K^-p$ scattering length $a_p$ extracted from different 
experiments/analyses by using Eq.~(\ref{apnlo}).}
\bigskip
\label{tab:ap}
\end{center}
\end{table}
Scattering lengths have also been extracted in Ref.~\cite{Borasoy:2006sr}
considering the meson-baryon interaction up to next-to-leading order in a 
coupled-channel chiral SU(3) unitary approach. In a combined fit to $K^-p$ 
induced reactions, different interaction kernels within the unitarization
scheme are considered and, using statistical criteria, non-linear errors on 
the values of $a_0$ and $a_1$ are provided (confidence regions in the complex
$a_I$ planes). Here, we use the extracted $a_p$ scattering length of the full 
approach of Ref.~\cite{Borasoy:2006sr}, $a_p=(-1.05+i\,0.75)$~fm. This value can 
be combined with the scattering length extracted from the latest experiment on 
kaonic atoms, SIDDHARTA~\cite{Bazzi:2011zj}. The average is shown in the last 
line of Tab.~\ref{tab:ap}, and we consider this number as the most reliable 
available value of $a_p$. In the following section, we will determine the 
restrictions on $a_0,\,a_1$ and also $A_{Kd}$ from these measurements.

\begin{figure}[h!]
\begin{center}
\includegraphics[width=9.cm]{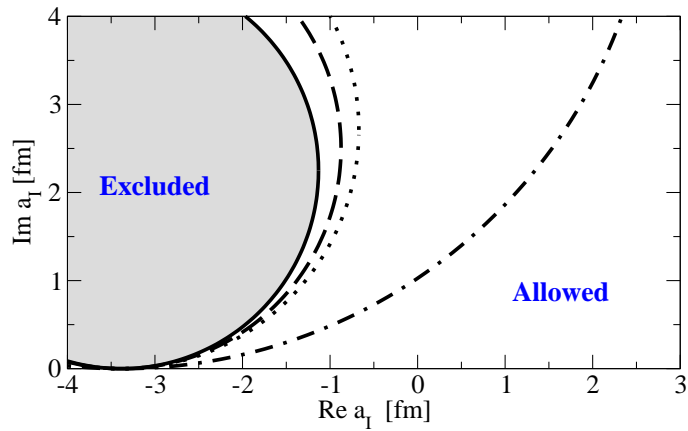}
\end{center}
\caption{Solid line: Restrictions on the values of the scattering  lengths
$a_0$ and $a_1$ set by the SIDDHARTA data~\cite{Bazzi:2011zj} combined with the
$K^-p$ scattering length obtained from scattering data~\cite{Borasoy:2006sr}. 
For comparison, we also display the restrictions from 
SIDDHARTA only (dashed line), DEAR~\cite{Beer} (dot-dashed line)  and
KpX~\cite{KEK} (dotted line).
}
\label{fig:circle}
\end{figure}

\medskip
\noindent
{\bf 3.} 
In Fig.~\ref{fig:circle}, we show the restrictions on the values of the
isoscalar and isovector scattering lengths corresponding to the
condition defined in Eq.~(\ref{circle}). We only show one quadrant of the whole
circle. The new determination of the $K^-p$ scattering length is less
restrictive than the one based on the DEAR results and not very different from
the one based on the older KpX measurement. This is another way of
demonstrating that now the determinations of the $K^-p$ scattering length from
kaonic hydrogen and from scattering data are consistent, see also Fig.~1 
of Ref.~\cite{Meissner:2006gx}.

\begin{figure}[t!]
\begin{center}
\includegraphics[width=9.6cm]{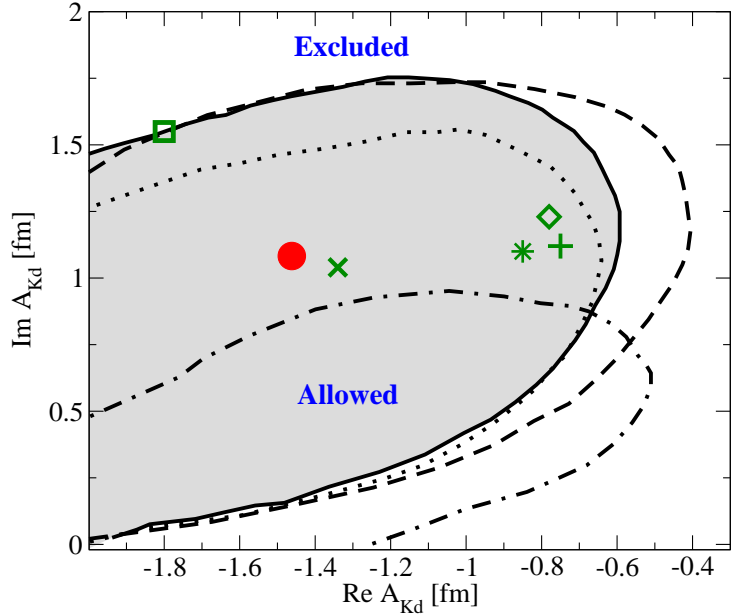}
\end{center}
\caption{The region in the $({\rm Re}\,A_{Kd}$,  ${\rm Im}\,A_{Kd})$--plane
where  solutions for $a_0$ and $a_1$ exist.  The grey area bounded by the solid
line is our central prediction using the average of the $K^-p$ scattering
length from SIDDHARTA \cite{Bazzi:2011zj} and Ref.~\cite{Borasoy:2006sr}. The
dashed, dot-dashed and dotted lines are generated using the experimental input
from SIDDHARTA, DEAR \cite{Beer} and KpX \cite{KEK}, respectively. The filled (red)
circle is the prediction for $A_{Kd}$ based on the $\bar KN$ S-wave scattering
lengths from Ref.~\cite{Borasoy:2006sr}, c.f. Eq.~(\ref{eq:AKdpred}). 
Older predictions for $A_{Kd}$ are from:
\cite{Torres:1986mr} (cross), \cite{Deloff:1999gc} (star: Faddeev equations, plus sign: FCA), 
\cite{Bahaoui:2003xb} (square), \cite{Grishina} (diamond).
}
\label{fig:area}
\end{figure}
In Fig.~\ref{fig:area}, we analyse the values the complex-valued kaon-deuteron
scattering length can take where solutions  for $a_0$ and $a_1$ exist at all.
For that, we have scanned the region  $-2~{\rm fm}< {\rm Re}\,A_{Kd}<0$ and  $0
< {\rm Im}\,A_{Kd}<2~{\rm fm}$  and tried to find solutions, using the input
data collected in Tab.~\ref{tab:ap}. Using our best determination of $a_p$ or
directly the one extracted from SIDDHARTA, the allowed region is  again
increased as compared to the one based on the DEAR data. For the central
values extracted from scattering data, $a_0 = (-1.64 + i\ 0.75)\,$fm and 
$a_1 = (-0.06 + i\ 0.57)\,$fm~\cite{Borasoy:2006sr}, we predict the kaon-deuteron
scattering length as:
\begin{equation}
A_{Kd}=(-1.46 + i\ 1.08)~~{\rm fm},
\label{eq:AKdpred}
\end{equation}
which is also displayed as a red circle in Fig.~\ref{fig:area}, together with other predictions
from the literature.
We note in this context the (somewhat model-dependent) limit on the $K^- d$ scattering length 
extracted from the $\bar K^0 d$ mass spectrum obtained from the reaction  
$pp \to d \bar K^0 K^+$ measured at the Cooler Synchrotron COSY at  J\"ulich, 
namely Im~$A_{Kd} \le 1.3\,$fm  and $|{\rm Re}~A_{Kd}| \le 1.3\,$fm
\cite{Sibirtsev:2004kk}. Our prediction for the real part in
Eq.~(\ref{eq:AKdpred}) is not in contradiction with this bound, because
the uncertainties in the determination of $a_0$ and $a_1$ given in 
Ref.~\cite{Borasoy:2006sr} lead to an approximate uncertainty of $\pm
0.35\,$fm in Re~$A_{Kd}$ (and a similar uncertainty in Im~$A_{Kd}$).

\begin{figure}[t!]
\begin{center}
\includegraphics[width=9.4cm]{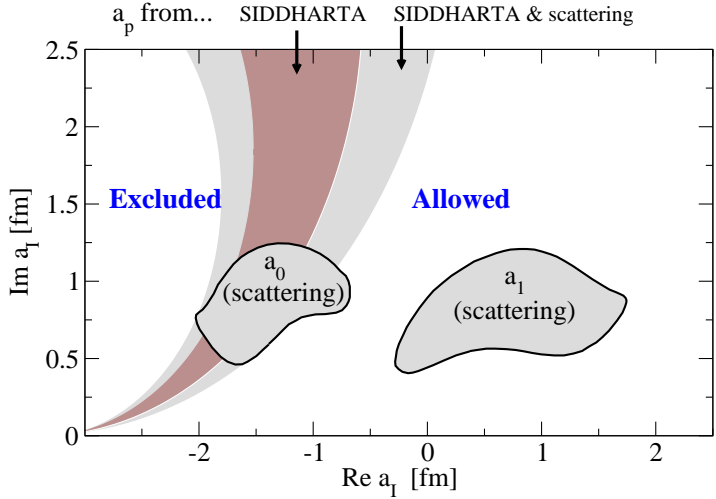}
\end{center}
\caption{Error estimates. Shaded areas with borders: allowed values from
scattering data,  result from Ref.~\cite{Borasoy:2006sr}. Dark shaded area:
Uncertainty of the border between allowed and excluded values,  using SIDDHARTA
uncertainties~\cite{Bazzi:2011zj}. Light shaded area: same, using combined
uncertainties from SIDDHARTA and scattering~\cite{Borasoy:2006sr} data.
}
\label{fig:circleerror}
\end{figure}

So far, we have worked with the central values of the input quantities.
To estimate the error of our predictions, the statistical and systematic errors of the SIDDHARTA
measurement have been added in quadrature, and the error of the extracted $a_p$
has been determined to be approximately $\Delta\,{\rm Re}\, a_p=0.10$~fm,
$\Delta\,{\rm Im}\, a_p=0.13$~fm. These values lead to an uncertainty of the
boundary between allowed and excluded values for $a_0$ and $a_1$, as indicated
with the dark shaded area in Fig.~\ref{fig:circleerror}. The uncertainties of
$a_p$ determined from scattering data~\cite{Borasoy:2006sr} are estimated to be
$\Delta \,{\rm Re}\, a_p\sim \Delta \,{\rm Im}\, a_p\sim 0.2$~fm. If the errors
from SIDDHARTA and scattering are added in quadrature, one obtains the light
shaded areas in Fig.~\ref{fig:circleerror}.

Note that scattering provides restrictions on $(a_0,a_1)$ that are not fully
encoded in the single value of $a_p$ from Ref.~\cite{Borasoy:2006sr}; indeed,
as indicated in Fig.~\ref{fig:circleerror}, scattering provides the irregularly
shaped areas for $a_0$ and $a_1$ which pin these values down much more
precisely. In any case, the figure clearly shows that there is no conflict
between scattering data and scattering lengths.

\begin{figure}[t!]
\begin{center}
\includegraphics[width=8cm]{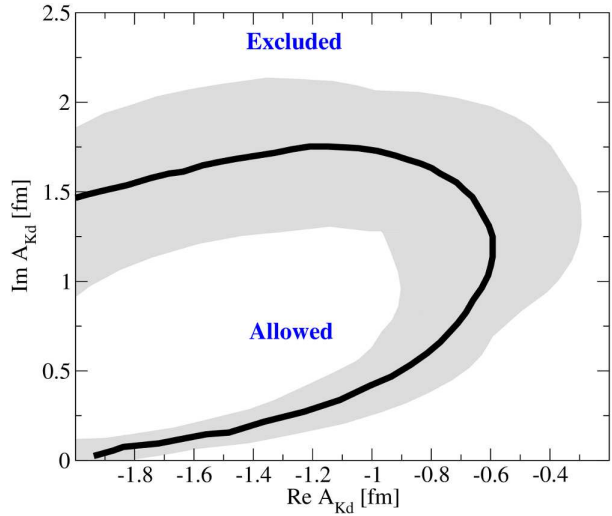}
\end{center}
\caption{Uncertainty of the boundary of allowed values for $A_{Kd}$. The
central value (solid line) corresponds to the average of $a_p$ from scattering
data and the SIDDHARTA value; uncertainty (shaded area) from the combined
errors of these two sources.
}
\label{fig:akderrors}
\end{figure}

The combined uncertainty of $a_p$ from scattering and kaonic atom data
translates into the uncertainty of the boundary of allowed values for the 
kaon--deuteron scattering length. This is indicated with the shaded area in
Fig.~\ref{fig:akderrors}. Additional uncertainty stems from the static approach
to $Kd$ scattering itself, i.e. Eq.~(\ref{eq:ratio-Kamalov}). Recoil
corrections can be sizeable and should be included in future
calculations. As for the genuine three-body term $\delta
\hat a_{Kd}$ in Eq.~(\ref{eq:ratio-Kamalov}), note that its imaginary part is
related to the total two-nucleon absorption rate of the $K^-$ which amounts to
$(1.22\pm 0.09)\%$~\cite{Veirs:1970fs}; assuming
${\rm Re}\,\delta \hat a_{kd}$ to be of similar size as ${\rm Im}\,\delta \hat
a_{Kd}$, the three-body force contributes only with a few percent to $A_{Kd}$. 

\medskip

\noindent
{\bf 4.}
In summary, we have reanalysed the predictions for the kaon-deuteron scattering
length in view of the new kaonic hydrogen experiment from SIDDHARTA. Based on
consistent solutions for input values of the $K^-p$ scattering length, we have
explored the allowed ranges for the isoscalar and isovector kaon-nucleon
scattering lengths and explored the range of the complex-valued kaon-deuteron
scattering length that is consistent with these values. 
In particular, the new SIDDHARTA measurement is shown to resolve inconsistencies
for $a_0$, $a_1$, and $A_{Kd}$ as they arose from the DEAR data.
A precise measurement
of the $K^-d$ scattering length from kaonic deuterium would therefore serve as
a stringent test of our understanding of the chiral QCD dynamics and is
urgently called for.

\bigskip

{\it Acknowledgments}

We thank Akaki Rusetsky for stimulating discussions. Partial financial support
from the EU Integrated Infrastructure Initiative HadronPhysics2 (contract
number 227431) and DFG (SFB/TR 16, ``Subnuclear Structure of Matter'') is
gratefully acknowledged.


\end{document}